\documentstyle[12pt,epsfig]{article}
\begin{document}
\pagestyle{empty}
\renewcommand{\thefootnote}{\fnsymbol{footnote}}
\def\lsim{\raise0.3ex\hbox{$<$\kern-0.75em\raise-1.1ex\hbox{$\sim$}}}
\def\gsim{\raise0.3ex\hbox{$>$\kern-0.75em\raise-1.1ex\hbox{$\sim$}}}
\def\noi{\noindent}
\def\nn{\nonumber}
\def\bea{\begin{eqnarray}}  \def\eea{\end{eqnarray}}
\def\beq{\begin{equation}}   \def\eeq{\end{equation}}
\def\beeq{\begin{eqnarray}} \def\eeeq{\end{eqnarray}}
\def\R{ {\rm R \kern -.31cm I \kern .15cm}}
\def\C{ {\rm C \kern -.15cm \vrule width.5pt \kern .12cm}}
\def\Z{ {\rm Z \kern -.27cm \angle \kern .02cm}}
\def\N{ {\rm N \kern -.26cm \vrule width.4pt \kern .10cm}}
\def\1{{\rm 1\mskip-4.5mu l} }
\def\lsim{\raise0.3ex\hbox{$<$\kern-0.75em\raise-1.1ex\hbox{$\sim$}}}
\def\gsim{\raise0.3ex\hbox{$>$\kern-0.75em\raise-1.1ex\hbox{$\sim$}}}
\def\sq{\hbox {\rlap{$\sqcap$}$\sqcup$}}
\vbox to 2 truecm {}
\centerline{\Large \bf Deep Inelastic Scattering Data and the}
\vskip 3 truemm  
\centerline{\Large \bf Problem of Saturation in Small-x Physics}

\vskip 1 truecm
\centerline{\bf A. Capella, E. G. Ferreiro, C. A. Salgado}
\centerline{Laboratoire de Physique Th\'eorique\footnote{Unit\'e Mixte de Recherche -
CNRS - UMR n$^{\circ}$ 8627}}  \centerline{Universit\'e de Paris XI, B\^atiment 210,
F-91405 Orsay Cedex, France}

\vskip 5 truemm
\centerline{\bf A. B. Kaidalov}
\centerline{ITEP, B. Cheremushkinskaya ulitsa 25}
\centerline{117259 Moscou, Russia}
\vskip 1 truecm 
\begin{abstract}
We investigate the role of unitarization effects in virtual photon-proton ($\gamma^*p$)
interactions at small $x$. The $q\bar{q}$-fluctuation of the initial photon is
se\-pa\-ra\-ted into a small distance and a large distance component and a model for the
unitarization of each component is proposed. The Born approximation for the small size
component is calculated using QCD perturbation theory. Reggeon diagram technique is used in
order to obtain a self-consistent scheme for both total $\gamma^*p$ cross section and
diffractive production. The model gives a good description of HERA data in the small-$x$
region, with a single Pomeron of intercept 1.2. 
\end{abstract}

\vskip 1 truecm

\noindent LPT Orsay 00-50 \par
\noindent June 2000 \par

\newpage
\pagestyle{plain}
\baselineskip=24 pt

\section{Introduction}
\hspace*{\parindent} The present work is an extension of our previous one \cite{1r} on
the investigation of unitarity effects in small-$x$ processes. It was found
experimentally at HERA that both the total cross section of a highly virtual photon,
$\sigma_{\gamma^*p}^{(tot)}$, and the cross section for its diffractive dissociation
have a fast increase with energy. This is related to a fast increase of densities of
quarks and gluons as the Bjorken variable $x$ decreases. The dynamics of such very
dense partonic systems is very interesting and has been studied by many authors both
in deep inelastic scattering (see ref. \cite{1paper1} for reviews and 
ref. \cite{2paper1} for some recent papers) 
and in high energy nuclear interactions \cite{3paper1}. 
Unitarity effects should stop the increase of
densities at extremely small $x$ and lead to a ``saturation'' of parton densities. It
is important to determine the region of $x$ and $Q^2$ where the effects of saturation
become important. \par

We study this problem using reggeon calculus \cite{2r} with a supercritical Pomeron
($\alpha_P(0) - 1 \equiv \Delta >0$) and the partonic picture of
$\gamma^*p$-interaction in QCD. In our previous paper \cite{1r} we used this approach
for the description of HERA data in the region $0 \leq Q^2 \leq Q_0^2$ $(Q_0^2 \sim
10\ {\rm GeV}^2)$, where the effects of unitarity are most important. It was shown
that, with a single Pomeron of intercept 1.2 and multipomeron exchanges (unitarity
effects), it is possible to obtain a self-consistent, simultaneous, description of
both the total $\gamma^*p$-cross section and diffractive production in high-energy
$\gamma^*p$-interactions. In such approach, it is convenient to consider the process
of $\gamma^*p$-interaction in the laboratory frame as an interaction of the
$q\bar{q}$-pair, produced by the photon, with the proton. We separated the
$q\bar{q}$-pair fluctuation into two components, ``aligned'' component with a strongly
asymmetric 
sharing of the momentum fraction $z$ between $q$ and $\bar{q}$, and the
rest (``symmetric'') component. Such a separation is important at large $Q^2$, where
the first component has a large transverse size, while the ``symmetric'' component
has a size $r \sim {1 \over Q}$ and thus has a small cross section $\sim 1/Q^2$ of
interaction with the target. Both components give a contribution to the
$\sigma_{\gamma^*p}^{(tot)}$ which behaves as $1/Q^2$ at large $Q^2$, but the
``aligned'' component gives the main contribution to the diffraction production cross
section. Triple Pomeron diagrams were also included in our model. \par

In this paper we propose a more direct separation of the two components of the
$q\bar{q}$-pair, which is valid also for small $Q^2$. The separation into a small
size ($S$) and a large size ($L$) components of the $q\bar{q}$ pair is now made in
terms of the transverse distance $r$ between $q$ and $\bar{q}$. The border value,
$r_0$, is treated as a free parameter - which turn out to be $r_0 \sim
0.2$~fm\footnote{This value agrees with the correlation length of nonperturbative
interactions observed in lattice calculations \protect{\cite{3r}}.}. \par

For the $S$-component, with $r \leq r_0$, we use the expression for the 
$\gamma^*p$
total cross-section obtained in perturbative QCD \cite{4r}\cite{5r}.

\beq
\label{1e}
\sigma_{\gamma^*p}^{(tot)T(L)}(s, Q^2) = \int_0^{r_0} d^2r \int_0^1 dz \left |
\psi^{T(L)}(r,z, Q) \right |^2 \ \sigma_S(r, s, Q^2) \quad , \eeq

\noi where $T$ and $L$ correspond to transverse and longitudinal polarizations of a
virtual photon, $\psi^{T(L)}(r, z)$ are the corresponding wave functions of the
$q\bar{q}$-pair: 

\beq
\label{wave1}
\left |\psi^{T}(r,z, Q) \right |^2={6\alpha_{e.m.}\over 4\pi^2}
\sum_q e_q^2\{ [z^2+(1-z^2)]\epsilon^2K_1^2(\epsilon r)+m_q^2K_0^2(
\epsilon r)\} ,
\eeq

\noi and
\beq
\label{wave2}
\left |\psi^{L}(r,z, Q) \right |^2={6\alpha_{e.m.}\over 4\pi^2}
\sum_q e_q^2\{ 4Q^2z^2(1-z)^2K_0^2(\epsilon r)\} ,
\eeq

\noi where $\epsilon^2=z(1-z)Q^2+m_q^2$. $K_0$ and $K_1$ are McDonald
functions. The sums are over quark flavors and we have taken 
$m_u=m_d=m_s\equiv m_S$. 

\noi
$\sigma_S(r, s, Q^2)$ is the total cross section for the interaction of
the $q\bar{q}$-pair with the proton. For the interaction of a small size dipole

\beq
\label{2e}
\sigma_S(r, s, Q^2) = r^2 \ f(s, Q^2) \quad .
 \eeq  

\noi As for the $L$ component, we use the same parametrizations introduced in ref.
\cite{1r} for the aligned component.

 \section{The model}
\hspace*{\parindent} 

We write the $\gamma^*p$ total cross section 

\beq
\label{3e}
\sigma_{\gamma^*p}^{(tot)}(s, Q^2) = {4 \pi^2 \alpha_{e.m} \over Q^2} \ F_2(x, Q^2)
\quad ,
 \eeq

\noi in the following form, using the impact parameter ($b$) representation

\beq
\label{4e}
\sigma_{\gamma^*p}^{(tot)}(s, Q^2) = 4 \int d^2b \ \sigma_{\gamma^*p}^{(tot)}(b, s, Q^2)
\eeq

\beq
\label{5e}
\sigma_{\gamma^*p}^{(tot)}(b, s, Q^2) = g_L^2(Q^2) \ 
\sigma_L^{(tot)}(b, s, Q^2) +
\sigma_S^{(tot)}(b, s, Q^2) \quad .\eeq

The function $g_L^2(Q^2)$ 
determines the coupling of the photon to the large size $q\bar{q}$
pair and is chosen in the form \cite{1r}

\beq
\label{6e}
g_L^2(Q^2) = {g_L^2(0) \over 1 + {Q^2 \over m_L^2}}  
 \eeq

\noi where $g_L^2(0)$ and $m_L^2$ are phenomenological parameters. \par

The cross section for the $L$-component, $\sigma_L^{(tot)}$, in the impact parameter
space, is chosen in the quasi-eikonal form \cite{7r}

\beq
\label{7e}
\sigma_L^{(tot)}(b, s, Q^2) = {1 - \exp (- C \ \chi_L(b,s, Q^2))
 \over 2C} \quad ,
\eeq

\beq
\label{8e}
\chi_L(s,b,Q^2) = {\chi_{L0}^P(b, \xi) \over 1 + a \ \chi_3(s,b,Q^2)} + 
\chi_{L0}^f (b,
\xi) \quad .
 \eeq

The eikonal functions $\chi_{L0}^k$ $(k = P, f)$ are written in a standard Regge form 

\beq
\label{9e}
\chi_{L0}^k(b, \xi) = {C_L^k \over \lambda_{0k}^L (\xi)} \exp \left (
\Delta_k \xi - {b^2 \over 4 \lambda_{0k}^L(\xi )} \right ) \quad ,
\eeq

\noi where

\beq
\label{10e}
\Delta_k = \alpha_k(0) - 1 \quad , \quad \xi = \ell n {s + Q^2 \over s_0 + Q^2} \quad ,
\quad \lambda_{0k}^L = R_{0kL}^2 + \alpha '_k \ \xi \quad .
\eeq

Here $\alpha_k(0)$ is the intercept of trajectory $k$ and $\alpha '_k$
its slope. The values of the radii $R_{0kL}^2$, based on ref. \cite{24p1},
are given in Table 1. The quantity $\xi$ is chosen in such a way as to behave 
as $\ell n {1 \over  x}$ for 
large $Q^2$ and as $\ell n {s \over s_0}$ for $Q^2 = 0$.

The coefficients $C_L^P$ and $C_L^f$ determine respectively the residues of the Pomeron
and $f$-reggeon exchanges in the $q\bar{q}$-proton interaction. The coefficient $C = 1.5$
takes into account the dissociation of a proton \cite{7r}. \par

We turn next to the denominator of eq. (\ref{8e}). The constant $a$ si given by $a =
{g_{pp}^P(0) \ r_{PPP}(0) \over 16 \pi}$, where $g_{pp}^P(0)$ is the proton-Pomeron
coupling and $r_{PPP}(0)$ is the triple Pomeron coupling, both at $t = 0$. The
function $\chi_3(b,s,Q^2)$ is given by eq. (\ref{29e}) of section 3. \par

With $a = 0$, the model described above is a standard quasi-eikonal model with Born terms
given by Pomeron plus $f$ exchanges. The denominator in eq. (\ref{8e}) cor\-res\-ponds to a
resummation of triple Pomeron branchings (the so-called fan diagrams). (For a full
discussion on the interpretation of this denominator see ref. \cite{1r}). Thus,
expressions (\ref{7e}) and (\ref{8e}) correspond to a sum of diagrams of the type shown in
Fig.~1. \par

We turn next to the $S$ component. In this case we put, in complete analogy with eqs.
(\ref{7e})-(\ref{9e})

\beq
\label{11e}
\sigma_S^{(tot)}(r, b, s, Q^2) = {1 - \exp \left ( - C \ \chi_S(r,b,s,Q^2) \right )
\over 2C} \quad ,
 \eeq

\beq
\label{12e}
\chi_S(r, b, s, Q^2) = {\chi_{S0}(r,b,s,Q^2) \over 1 + a \ \chi_3(b,s,Q^2)} \quad , 
 \eeq

\beq
\label{13e}
\chi_{S0}(r, b, \xi) = {C_S^P \ r^2 \over \lambda_{0P}^S(\xi)} \exp \left (
\Delta_P\xi - {b^2 \over 4 \lambda_{0P}^S(\xi )} \right )\quad , 
 \eeq

\noi with $\lambda_{0P}^S = R_{0PS}^2 + \alpha '_P \xi$. \par

Note that the contribution of the $f$-exchange to the $S$ component is very small and has
been neglected \cite{1r}. The condition (\ref{2e}), valid for fixed $s$ and $Q^2$ as $r
\to 0$, is a property of the single Pomeron exchange. Thus a factor $r^2$ has been
introduced in eq. (\ref{13e}). \par

Finally $\sigma_S(r, s, Q^2)$ in eq. (\ref{1e}) is obtained from $\sigma_S(r, b,s, Q^2)$,
defined by eqs. (\ref{11e}) to (\ref{13e}), as (see eq. (\ref{4e}))

\beq
\label{14e}
\sigma_S(r,s, Q^2) = 4 \int d^2b \ \sigma_S(r, b,s,Q^2) \quad . 
\eeq  

\noi Inserting this expression in eq. (\ref{1e}) we obtain the transverse and longitudinal
contributions of the $S$-component to the total $\gamma^*p$ cross-section.

\section{Diffractive production}
\hspace*{\parindent} Following ref. \cite{1r} we express the total diffractive
dissociation cross-section of a virtual photon as a sum of three terms

\beq
\label{15e}
\sigma_{\gamma^*p}^{(diff)} = \sum_{i=L,S} \sigma_i^{(0)} + \sigma_{PPP} 
 \eeq

\noi where

\beq
\label{16e}
\sigma_L^{(0)} = 4 g_L^2(Q^2) \int \left ( \sigma_L^{(tot)}(b,s, Q^2) \right )^2 d^2b
\quad ,
 \eeq 

\beq
\label{17e}
\sigma_S^{(0)T,L} = 4 \int d^2b \int_0^{r_0} d^2r \int_0^1 dz \left | \psi^{T,L}(z, r)
\right |^2 \left (  \sigma_S^{tot}(r, b, s, Q^2) \right )^2 
 \eeq 

\bea
\label{18e}
&&\sigma_{PPP} = 2 g_L^2(Q^2) \int \chi^L_{PPP}(b,s,Q^2)  \ e^{- 2C
\chi_L(b,s,Q^2)} \
d^2b \nonumber \\
&&+ 2 \int d^2b \int_0^{r_0^2} d^2r \int_0^1 dz \sum_{T,L} \left | \psi^{T,L}(z,
r) \right |^2  \chi_{PPP}^S (b, s, Q^2) \ e^{-2C\chi_S(r,b,s, Q^2)}  . \eea

\noi Here 

\beq
\chi_{PPP}^L(b,s, Q^2) = a \ \chi_L^P(b,s,Q^2) \ \chi_3(b,s,Q^2)
\label{19e}
\eeq

\noi and

\beq
\chi_{PPP}^S(r, b,s, Q^2) = a \ \chi_S(r, b,s,Q^2) \ \chi_3(b,s,Q^2)
\label{20e}
\eeq

\noi where $\chi_L^P(b,s,Q^2)$ is given by the first term of eq. (\ref{8e}) and
$\chi_3(b,s,Q^2)$ is defined by eq. (\ref{29e}). Using this expression, we see that, to
first order in $a$, $\sigma_{PPP}$ consists of the sum of a triple Pomeron ($PPP$) term
plus a $PfP$ one. We call this sum triple Pomeron, although the second one is an
interference term. For the total diffractive production cross-section,
that includes the diffraction dissociation of a proton, eqs. 
(\ref{16e})-(\ref{18e}) must be multiplied by the same factor $C$=1.5 of the
total $\gamma^*p$ cross-section.\par

At HERA, differential diffractive cross sections are given as a function of $\beta = {Q^2
\over M^2 + Q^2}$, where $M$ is the mass of the diffractively produced system, or of $x_P
= x/\beta$. They are usually integrated over $t$, and the function $F_{2D}^{(3)}$ is
introduced

\beq
\label{21e}
x_P \ F_{2D}^{(3)} = {Q^2 \over 4 \pi^2 \alpha_{e.m.}} \int x_P \ {d\sigma \over dx_Pdt} dt
\quad . \eeq  

\noi In our model, this function can be written as a sum of three terms

\beq
\label{22e}
F_{2D}^{(3)} = \left ( \sum_{i=L,S} F_{2Di}^{(3)} (x, Q^2,\beta ) + F_{2DPPP}^{(3)}(x, Q^2,
\beta ) \right ) \quad .\eeq

\noi Here


\beq
\label{23e}
x_P\ F_{2DL}^{(3)} = {Q^2 g_L^2(Q^2)\over 4 \pi
\alpha_{e.m.}}\ {\sigma_L^{(0)}\over \sigma_L^{(0)B}} 
\sum_{i,k=P,f}\int d^2b 
\chi_L^i\chi_L^k
{\widetilde{\beta}^{\Delta_i+\Delta_k-\Delta_f} (1 -
\beta)^{n_P(Q^2)} \over \int_{\beta_{min}}^{\beta_{max}} {d\beta \over \beta}
\widetilde{\beta}^{\Delta_i+\Delta_k-\Delta_f}(1 - \beta )^{n_P(Q^2)}} \eeq

\noi and

\beq
\label{24e}
x_P\ F_{2DS}^{(3)} = {Q^2 \over 4 \pi
\alpha_{e.m.}}\left (  \sigma_S^{(0)T} {\widetilde{\beta}^3 (1 -
2\beta)^2 \over \int_{\beta_{min}}^{\beta_{max}} {d\beta \over \beta}
\widetilde{\beta}^3(1 - 2\beta )^2} +  \sigma_S^{(0)L} {\widetilde{\beta}^3 (1 -
\beta) \over \int_{\beta_{min}}^{\beta_{max}} {d\beta \over \beta}
\widetilde{\beta}^3(1 - \beta )} \right )\eeq

\noi where $\widetilde{\beta} = {Q^2 + s_0 \over Q^2 + M^2}$, $\beta_{min} = {x \over
x_P^{max}} = 10 x$ and $\beta_{max} = {Q^2 \over M^2_{min} + Q^2}$ with 
$M_{min}^2 = 4 m_{\pi}^2$. In eq. 
(\ref{23e}) $\sigma_L^{(0)B}$ corresponds to eq. (\ref{16e}) keeping only
the linear term in $\sigma_L^{(tot)}$ and $\chi_L^{P(f)}$ is the 
contribution of the $P(f)$ in eq. (\ref{8e}).

The $\beta$-dependence of the $S$-component has been taken from the QCD results of ref.
\cite{6r}.
The $\beta$-dependence of the $L$-component was chosen according to ref. 
\cite{8r} and 

\beq
\label{25e}
n_P(Q^2) = - {1 \over 2} + {3 \over 2} \left ( {Q^2 \over c + Q^2} \right ) \quad , 
\eeq

\noi with $c = 3.5$~GeV$^2$. \par

The triple-Pomeron (i.e. $PPP$ plus $PfP$) contribution, $F_{2DPPP}^{(3)}(x,Q^2,\beta )$,
is given by

\beq
\label{26e}
x_P \ F_{2DPPP}^{(3)}(x, Q^2, \beta ) = x_P \ F_{2DPPP}^{(3)B}(x, Q^2, \beta ) \ 
{\sigma_{PPP} \over \sigma_{PPP}^B} \quad , \eeq 

\noi where $\sigma_{PPP}$ is given by eq. (\ref{18e}), its Born term, $\sigma_{PPP}^B$, by
the same equation with $C = 0$, and

\bea
\label{27e}
&&x_P\ F_{2DPPP}^{(3)B}(x, Q^2, \beta) = {Q^2 \over 4 \pi^2 \alpha_{e.m.}} 2a \int
d^2b\  \chi_3(b, s, Q^2, \beta) \times \nn \\
&& \left \{ g_L^2(Q^2) \ \chi_L^P(b,s,Q^2) + \sum_{T,L}
\int_0^{r_0} d^2r \int_0^1 dz \left | \psi^{T,L}(r,z) \right |^2 \ \chi_S(r,b,s,Q^2)
\right \}  . \eea

\noi Here

\beq
\label{28e}
\chi_3(s,b,Q^2,\beta ) = \sum_{k=P,f} \gamma_k \exp \left ( - {b^2 \over 4 \lambda_k\left (
{\widetilde{\beta} \over \tilde{x}}\right )} \right ) \left ( {\widetilde{\beta} \over
\widetilde{x}} \right )^{\Delta_k} {(1 - \beta )^{n_P(Q^2) + 4} \over \lambda_k\left (
{\widetilde{\beta} \over \tilde{x}}\right )} \eeq

\noi where $\gamma_P = 1$, $\gamma_f$ determines the strength of the $PfP$-contribution
relative to the $PPP$ one, and $\lambda_k = R_{1k}^2 + \alpha'_k \ell n \left (
{\widetilde{\beta} \over \tilde{x}}\right )$. The function $\chi_3(s, b, Q^2)$, which
enters in eqs. (\ref{8e}), (\ref{12e}), (\ref{19e}) and (\ref{20e}) is given by

\beq
\label{29e}
\chi_3(s,b,Q^2) = \int_{\beta_{min}}^{\beta^{max}} {d \beta \over \beta} \ \chi_3(s,b,Q^2,
\beta ) \quad . \eeq

\noi Since the triple Pomeron formula is not
valid for low masses, we use here $M_{min} =
1$~GeV.     

\section{Comparison with experiments}

The model was used to perform a fit of the data on structure function $F_2(x,
Q^2)$ and diffractive structure function $F_{2D}^{(3)}(x,Q^2,\beta)$, in the
region of small $x$ ($x<10^{-2}$) and $Q^2<10$~GeV$^2$\footnote{Actually, 
for $F_2$ only data with
$Q^2 \leq 3.5$ GeV$^2$ were included in the fit.}. In Table 1, the full 
list of parameters (fitted and fixed) is given. 

In Fig. 2, the results for $F_2(x,Q^2)$ are given as a function of $x$ for 
different values of $Q^2$. The description of the data is good. $S$ and $L$
contributions are shown separately in order to see the different behaviors. 
The $S$ contribution is almost negligible for very small $Q^2$, becoming 
comparable to the $L$ one at larger $Q^2$ values. 

In Figs. 3 to 6 we compare the model with the experimental data on
diffraction. In order to do such a comparison, it is necessary to take into 
account that different experiments use slightly different definitions
of diffractive events. In this way we have multiplied eq. 
(\ref{16e})-(\ref{18e}) by a factor $C_{diff}$=1.1 in order to compare with
data from H1 experiment and $C_{diff}$=1.3 for ZEUS\footnote{Notice that we
have taken these values as constant for each experiment, though they could
also depend on $M$. This would improve the agreement in Fig. 5.1.}. 
With these factors we 
take into account the different cuts in the mass of the diffractively 
dissociated proton (larger in the case of ZEUS).
As in the previous case, we plotted $L$, $S$ and $PPP$ contributions separately.
In Fig. 3, we show our results for the $\beta-$dependence of
$x_PF_{2D}^{(3)}$ for $x_P$=0.003 and for two values of $Q^2$. In Fig. 4,
the results are given as a function of $x_P$ for different values of
$\beta$ and $Q^2$. For the highest values of $Q^2$, only comparison with 
$\beta$=0.4 and $\beta$=0.65 are given. For smaller values of $\beta$,
QCD evolution becomes 
important. In Figs. 5, the energy dependence of the diffraction
cross-section is shown for different values of $M$ and $Q^2$. In Fig. 6, the
$M^2-$dependence of the model on diffractive dissociation in photoproduction  
is compared with HERA data for two different energies. Only data with $M^2<$
100 GeV$^2$ are shown for comparison. For larger values, the effect
of the non-diffractive $RRP$ contribution (not included in the model) is
expected to be large. 

 \section{Conclusions} 
\hspace*{\parindent} We have introduced a model of the eikonal type to describe 
total and
diffractive $\gamma^*p$ interactions. The $\gamma^*p$ interaction is viewed as 
that of a
$q\bar{q}$ pair, produced by the virtual photon, with the proton. The 
$\gamma^*p$ total
cross-section is separated into two components~: large size ($L$) for $r > r_0$ 
and small
size ($S$) for $r < r_0$, where $r$ is the transverse 
distance between $q$ and $\bar{q}$.
The value of $r_0$ - treated as a free parameter -
turns out to be $r_0 \sim 0.2$~fm. 
For the $L$-component, all the
$Q^2$-dependence is given by the coupling of $\gamma^*$ to the 
large size $q\bar{q}$
pair - which is taken as $1/Q^2$ at large $Q^2$ (eq. (\ref{6e})). For 
the $S$-component, the
$Q^2$-dependence is given by the wave function of the $q\bar{q}$ pair
(eqs. (\ref{wave1}, \ref{wave2})), computed in
perturbative QCD.
At large $Q^2$, $r^2\sim 1/Q^2$, and the unitarity corrections
of the $S$ component are higher twist, whereas those of the $L$ don't
depend on $Q^2$. \par

A good description of the small $x$ data is obtained both for $F_2$ and 
diffractive
production, in a broad region of $Q^2$ ($0 \leq Q^2 \ \lsim\ 10$~GeV$^2$), 
with a single
Pomeron of intercept $\alpha_P(0) = 1.2$. For larger values of $Q^2$, QCD 
evolution
becomes important. In particular it will give rise to a behavior $F_2 
\sim x^{-\Delta_P}$
with $\Delta_P$ significantly larger than 0.2 at large $Q^2$ \cite{meri}. 
For diffraction, this evolution 
has rather small effects at intermediate values of $\beta$ \cite{kdiff}.
This allows us to use our model, in this case,
without QCD evolution, up to rather
large $Q^2$ and moderate $\beta$.
\par

In the region $0 \leq Q^2 \leq 10$~GeV$^2$ the unitarity effects are 
very important and
produce a significant decrease of the effective Pomeron intercept 
$\alpha_P(0) = 1 +
\Delta_P$ with decreasing $Q^2$. This decrease is controlled by the 
strength of the
unitarity corrections. This, in turn, is controlled by the ratio
$\sigma^{(diff)}/\sigma^{(tot)}$ and its dependence on $Q^2$. 
Hence the importance of
describing both total cross-sections and diffractive production. 
In our case, $\chi_L > \chi_S$ and
the unitarity corrections are more important in the $L$ component
than in the $S$. Moreover, these corrections are higher twist at large
$Q^2$ in the second case. 
This 
is more clearly seen in the diffraction, where the
$S$ contribution to 
$x_PF_{2D}^{(3)}(x,Q^2,\beta)$ is much smaller than the $L$ one for
all but the larger $\beta$ values.
\par

An important result of our analysis is not only the fact that we can 
describe the 
data on both structure function and diffractive production in a broad 
region of $Q^2$ with
a single Pomeron, but also that we can describe diffractive production 
at $Q^2 = 0$ and at
intermediate $Q^2$ using the same value of the triple Pomeron coupling 
(which appears in
our parameter $a$). \par

Finally, we would like to discuss the large $\ell n(1/x)$ limit 
of the total $\gamma^*p$ cross-section in our model. 
The $\sigma^{(tot)}_L(b,s,Q^2)$, given by eqs. (\ref{7e})-(\ref{9e}),
tend to saturate fastly with increasing $s$ to the value $1/2C$,
due to the large $\chi_L(s,b,Q^2)$.
The situation is different for the $S$ component. Let's forget for the moment
about the triple pomeron contribution, i.e. consider the case $a$=0.
As we have said, unitarity corrections are much smaller in the $S$ component,
so, saturation will take place at much bigger energies, when the 
$\exp (\xi \Delta_P)$
term gets large enough. For such energies, cross section in the
small impact parameter will saturate to a $Q^2$-independent value and 
$F_2(x,Q^2)\sim Q^2$. This is the usual picture in perturbative QCD 
\cite{1paper1}-\cite{3paper1}.
However, by including the nonperturbative (large distance) $PPP$ terms 
($a\neq$ 0) we obtain a different behavior.
Indeed, the large $\exp (\xi 
\Delta_P)$ factors in
the numerator and denominator of eq. (\ref{12e}) cancel with each 
other, and we have
$\sigma_{\gamma^*p}^{(tot)} \sim {1 \over Q^2} f (\ell n Q^2$).
Thus, the $1/Q^2$ smallness of the $\gamma^*p$ cross-section is 
maintained in the limit
$x \to 0$. 

\noi {\bf Acknowledgments}

It is a pleasure to thank K. Boreskov, O. Kancheli, G. Korchemski, U. Maor and C
. Merino
for discussions. This work is partially supported by NATO grant OUTR.LG 971390.
E. G. F. and C. A. S. thank Ministerio de Educaci\'on y
Cultura of Spain for financial support.

\newpage

\section*{Figure captions}

\vskip 0.5cm

\noindent{\bf Figure 1.}
A generic reggeon diagram of our model. It contains the s-channel iteration
of Pomeron and $f$ exchanges, triple Pomeron ($PPP+PfP$) diagrams, as well as
multiple t-channel branchings of the Pomeron of the fan-diagram type.

\noindent{\bf Figure 2.}
$F_2(x,Q^2)$ as a function of $x$ for different values of $Q^2$ compared
with experimental data from H1 1995 \cite{h195} (open squares), ZEUS 1995
\cite{zeus1995} (black circles), E665 \cite{e665} (black triangles) (notice
that the corresponding $Q^2$ values of these data are slightly different)
and ZEUS BPT97 \cite{zeus1997} (open circles). Dotted curve corresponds
to the $L$ contribution, dashed one to the 
$S$ contribution and solid one to the total
$F_2(x,Q^2)$ given by the model.

\noindent{\bf Figure 3.}
$x_PF_{2D}^{(3)}$ as a function of $\beta$ for fixed $x_P=0.003$ and 
for $Q^2$=4.5
GeV$^2$ and $Q^2$=7.5 GeV$^2$. Experimental data are from \cite{h1diff}.
Dotted lines correspond to PPP contribution, dashed ones to L term and
dotted-dashed to S one.

\noindent{\bf Figure 4.1.}
$x_PF_{2D}^{(3)}$ as a function of $x_P$ for $Q^2$=4.5 and 7.5
GeV$^2$ and fixed $\beta$=0.04, 0.1, 0.2, 0.4, 0.65 and 0.9. 
The curves correspond to 
the convention of Fig. 3. Experimental data are from \cite{h1diff}.

\noindent{\bf Figure 4.2.}
$x_PF_{2D}^{(3)}$ as a function of $x_P$ for $Q^2$=9, 12 and  
18 GeV$^2$ and fixed 
$\beta$=0.4 and 0.65. The curves correspond to 
the convention of Fig. 3. Experimental data are from \cite{h1diff}.

\noindent{\bf Figure 5.1.}
Energy dependence of diffractive cross section for $M$=2, 5 and 11 GeV
and $Q^2$=8 and 14 GeV$^2$. Experimental data are from
\cite{zeusdiff}. The curves correspond to the convention of Fig. 3.

\noindent{\bf Figure 5.2.}
Energy dependence of diffractive cross section for different mass intervals and
for low-$Q^2$ compared with experimental data from
\cite{zeusdifflowq}. The curves correspond to the convention of Fig. 3.

\noindent{\bf Figure 6.}
Diffractive photoproduction cross-section for $W$=187 and 231 GeV as
a function of $M^2$ from \cite{photo} compared with our model. The 
curves follow the
same convention as in Fig. 3.

\newpage

\begin{center}
{\bf Table 1}
\vskip 1cm
\begin{tabular}{|c|c|c|c|} \hline
\multicolumn{2}{|c|} {Fixed Parameters} & \multicolumn{2}{c|} {Fitted 
Parameters} \\ \hline 
$\Delta_P$ & 0.2 & $g_L^2(0)$ & 4.56$\times 10^{-3}$\\ \hline 
$\Delta_f$ & -0.3 & $C_L^f$ & 1.97 GeV$^{-2}$\\ \hline
$\alpha^{\prime}_P$ & 0.25 GeV$^{-2}$ & $C_L^P$ & 0.56 GeV$^{-2}$\\ \hline 
$\alpha^{\prime}_f$ & 0.9 GeV$^{-2}$ & $s_0$ & 0.79 GeV$^2$\\ \hline
$R_{0kL}^2$ & 3 GeV$^{-2}$ & $a$ & 4.63$\times 10^{-2}$ GeV$^{-2}$\\ \hline  
$R_{0PS}^2$ & 2 GeV$^{-2}$ & $m_L^2$ & 0.59 GeV$^2$\\ \hline  
$R_{1k}^2$ & 2.2 GeV$^{-2}$ & $C_S$ & 0.18 \\ \hline 
$\gamma_f$ & 8  & $r_0$ & 1.06 GeV$^{-1}$\\ \hline 
$C$ & 1.5 & $m_S^2$ & 0.15 GeV$^2$\\ \hline 
\end{tabular}
\end{center}

\newpage

\centerline{\bf Figure 1}
\vspace{1cm}

\begin{center}
\hspace{-1.2cm}\epsfig{file=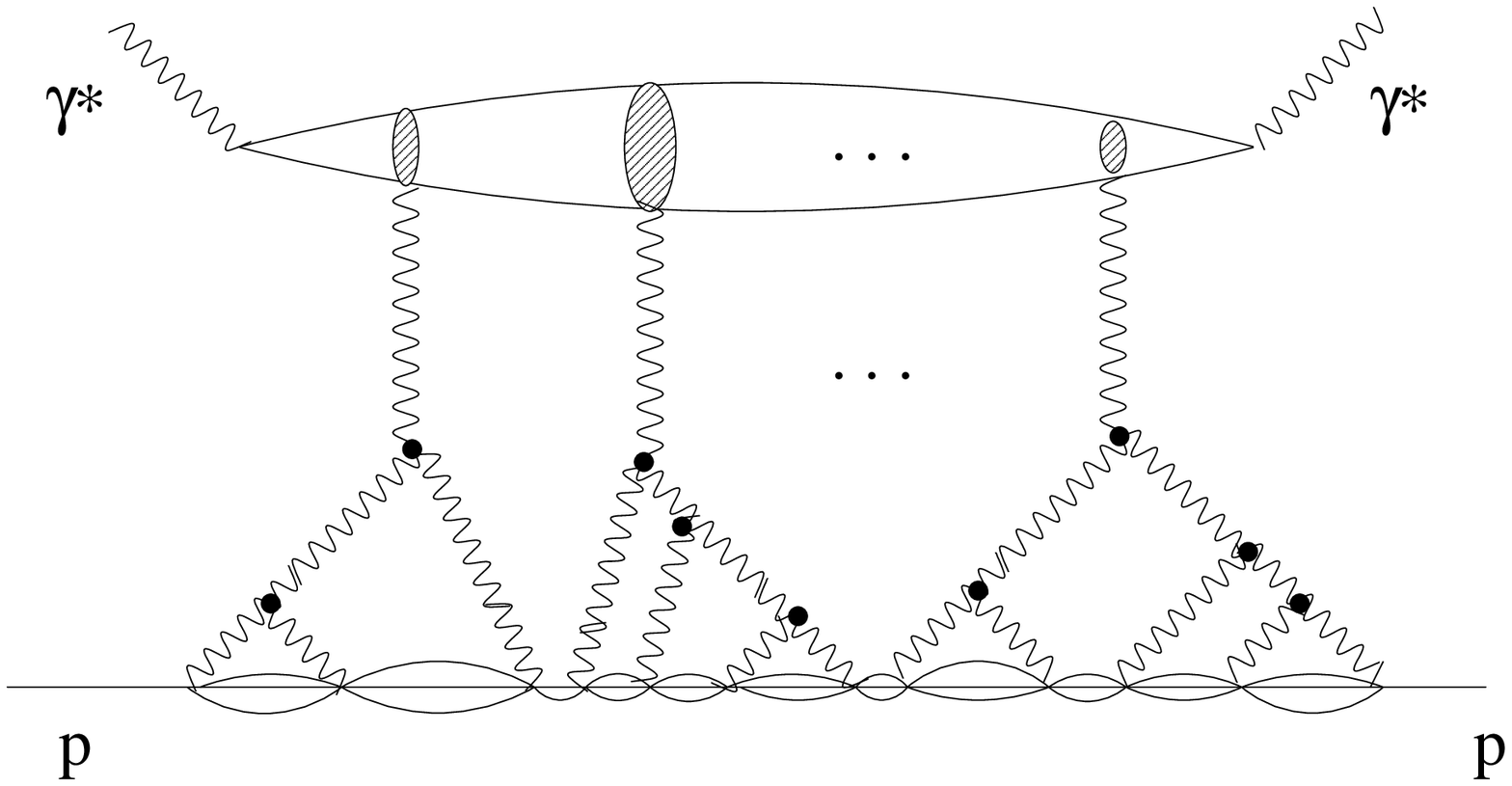,width=13.cm}
\end{center}

\newpage

\centerline{\bf Figure 2}
\vspace{1cm}

\begin{center}
\hspace{-1.2cm}\epsfig{file=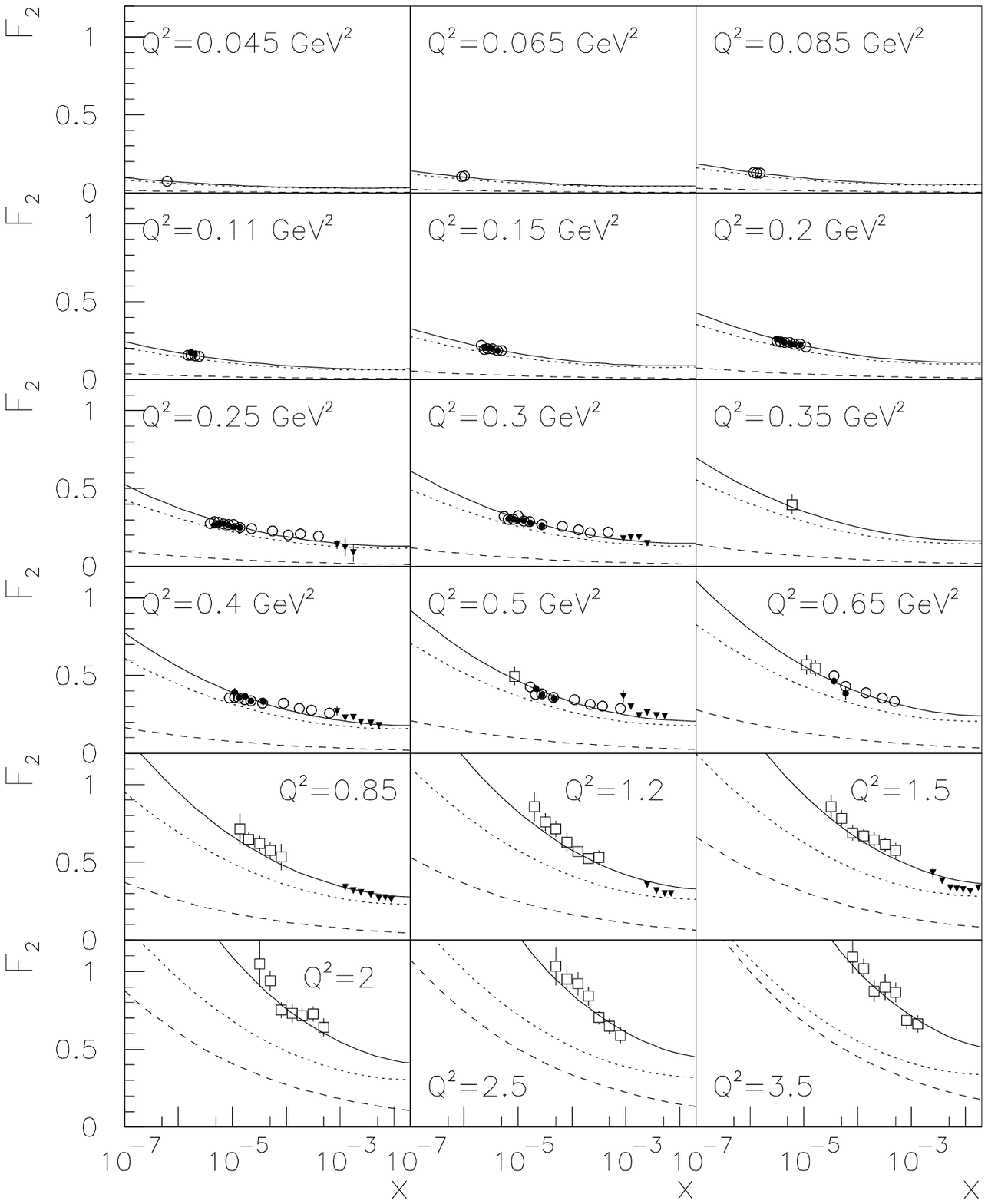,width=14.cm}
\end{center}

\newpage

\centerline{\bf Figure 3}
\vspace{1cm}

\begin{center}
\hspace{-1.2cm}\epsfig{file=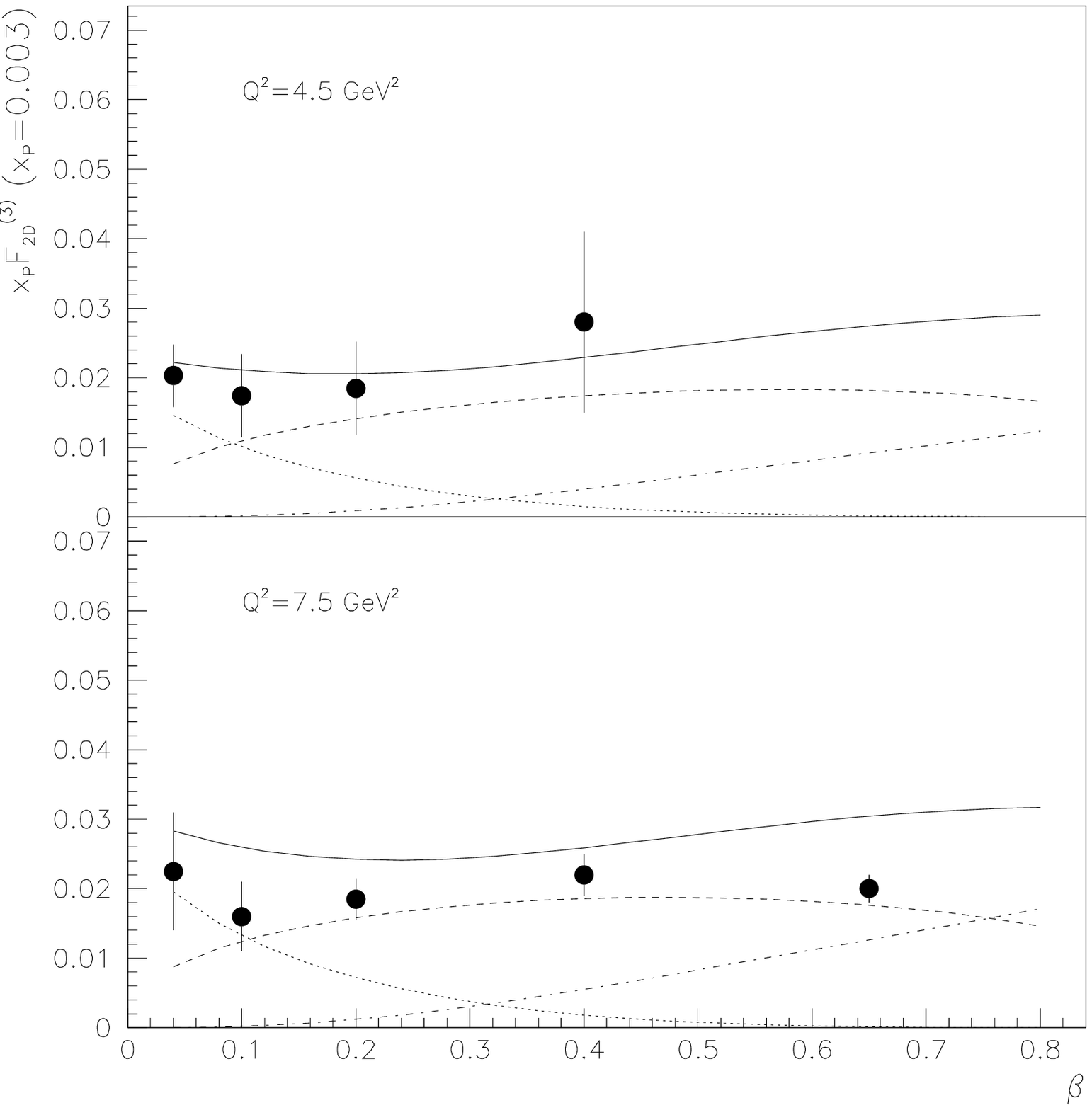,width=16.cm}
\end{center}

\newpage

\centerline{\bf Figure 4.1}
\vspace{1cm}

\begin{center}
\hspace{-1.2cm}\epsfig{file=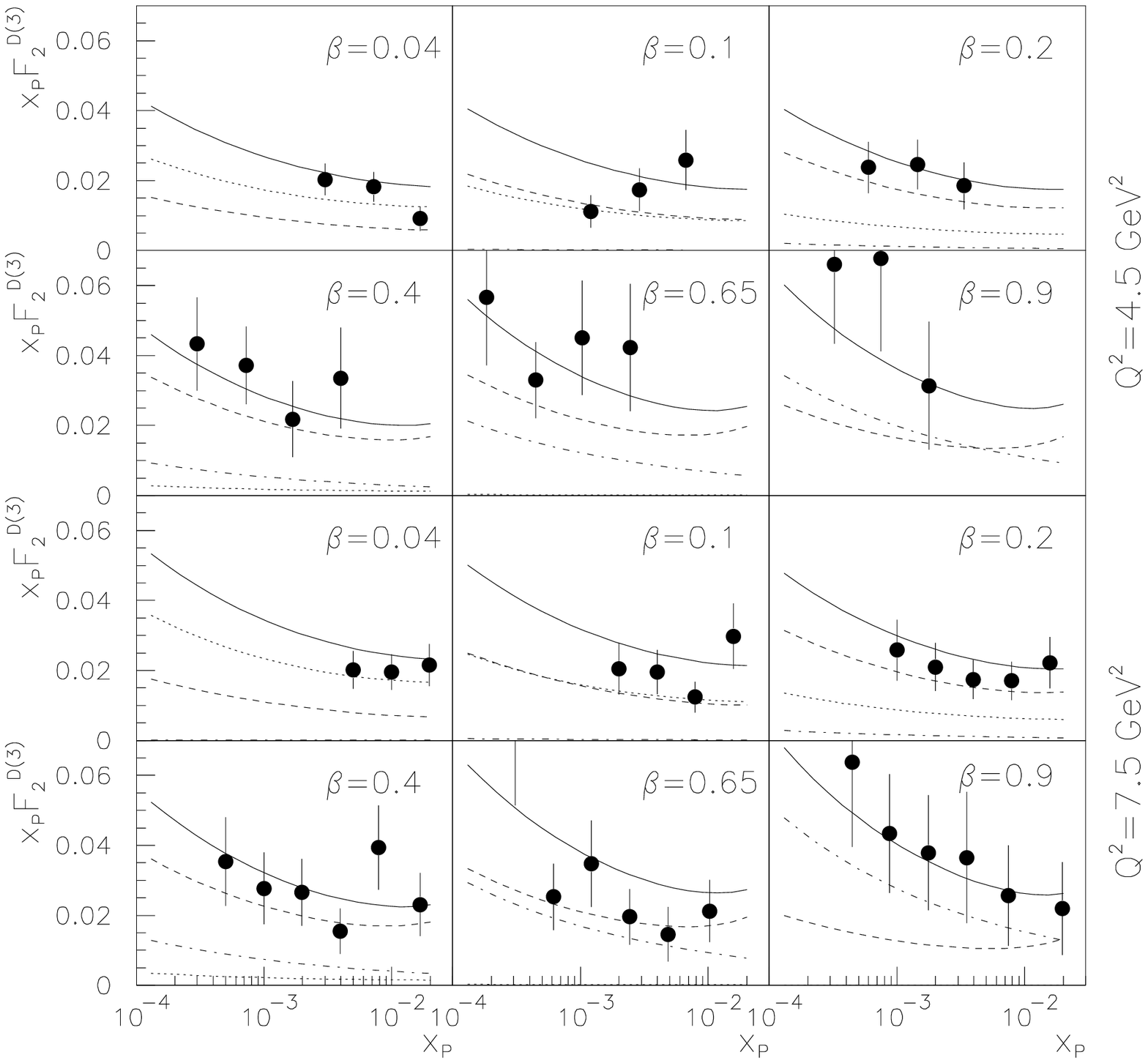,width=16.cm}
\end{center}

\newpage

\centerline{\bf Figure 4.2}
\vspace{1cm}

\begin{center}
\hspace{-1.2cm}\epsfig{file=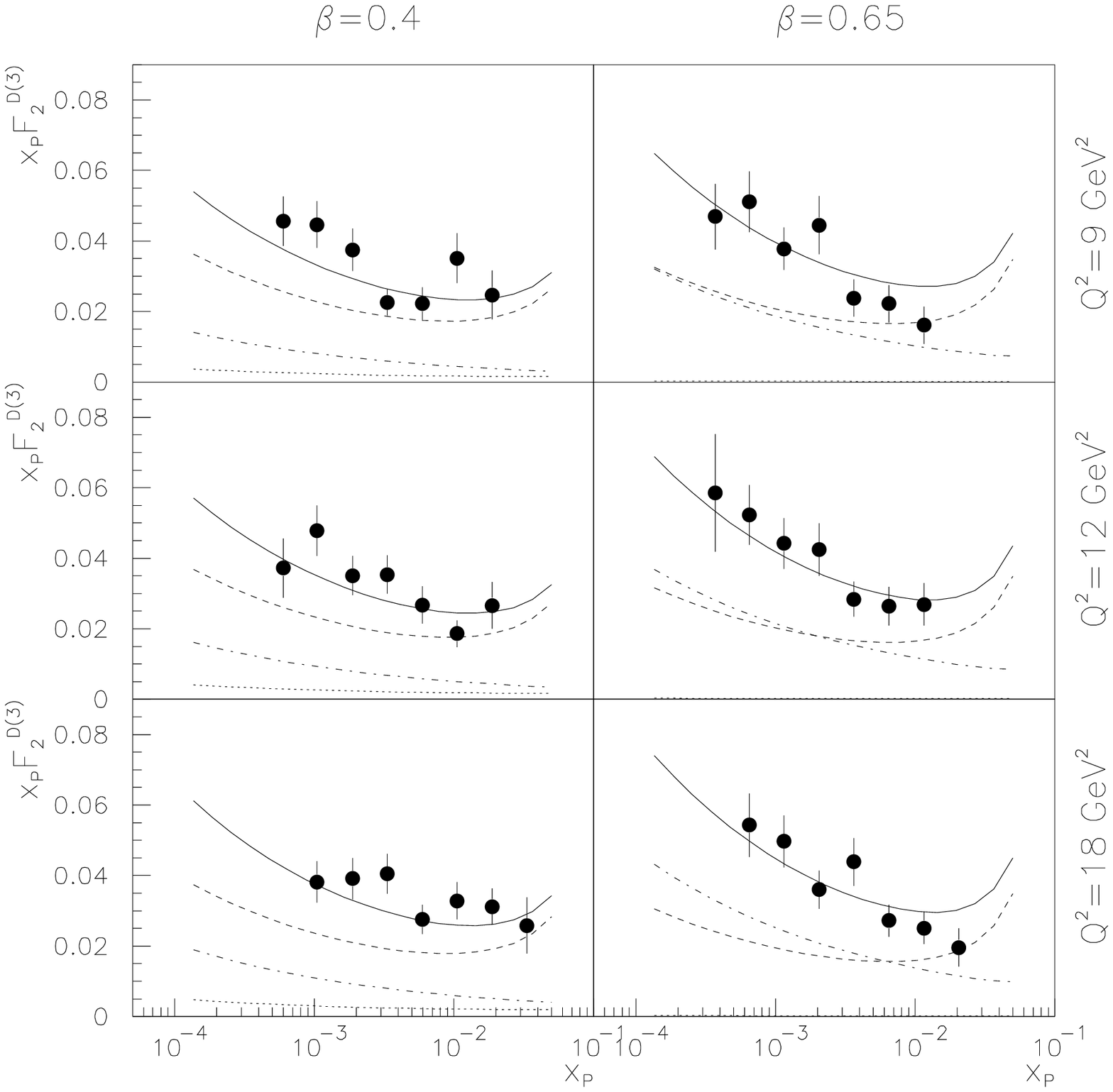,width=16.cm}
\end{center}

\newpage

\centerline{\bf Figure 5.1}
\vspace{1cm}

\begin{center}
\hspace{-1.2cm}\epsfig{file=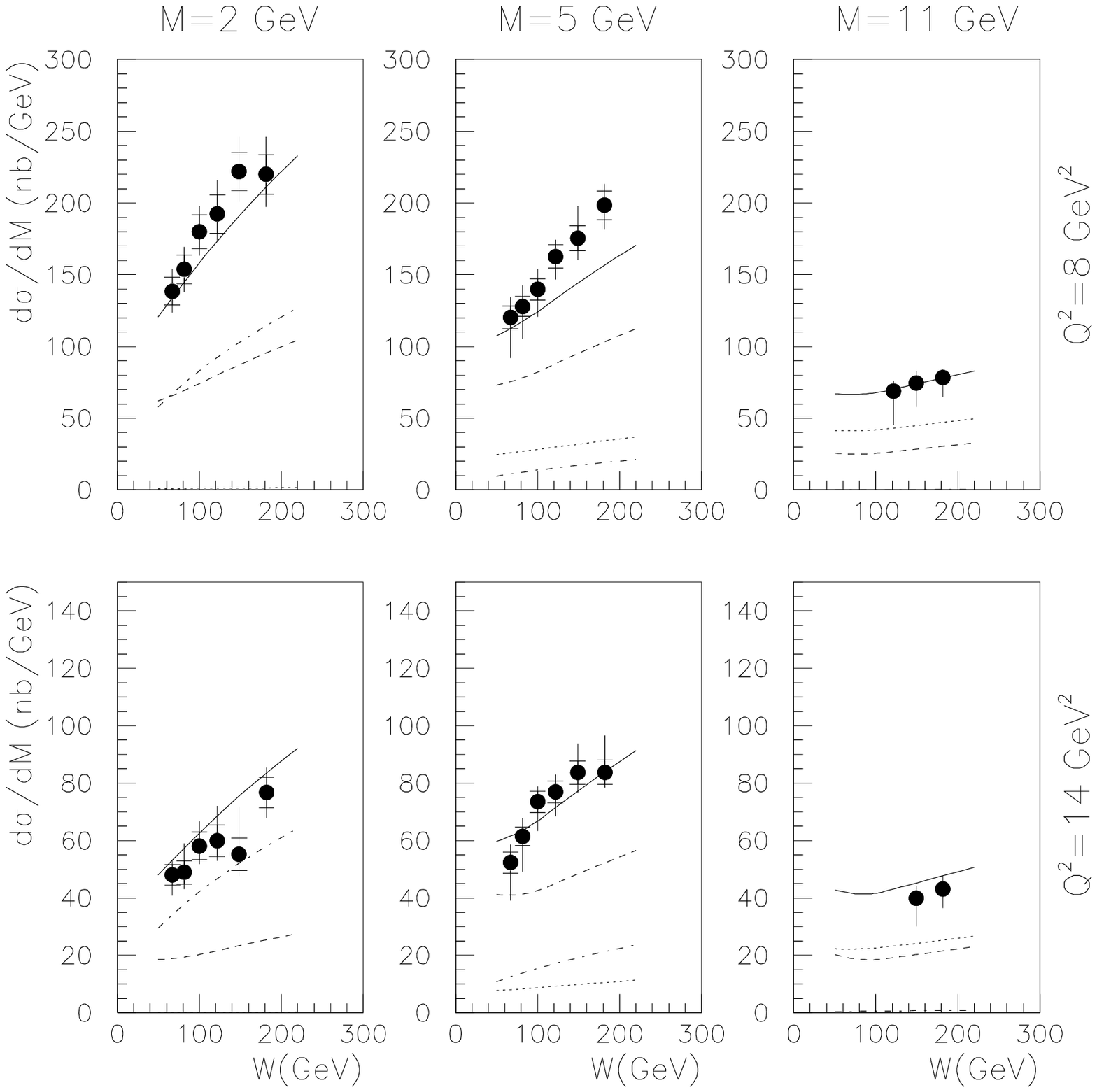,width=16.cm}
\end{center}

\newpage

\centerline{\bf Figure 5.2}
\vspace{1cm}

\begin{center}
\hspace{-1.2cm}\epsfig{file=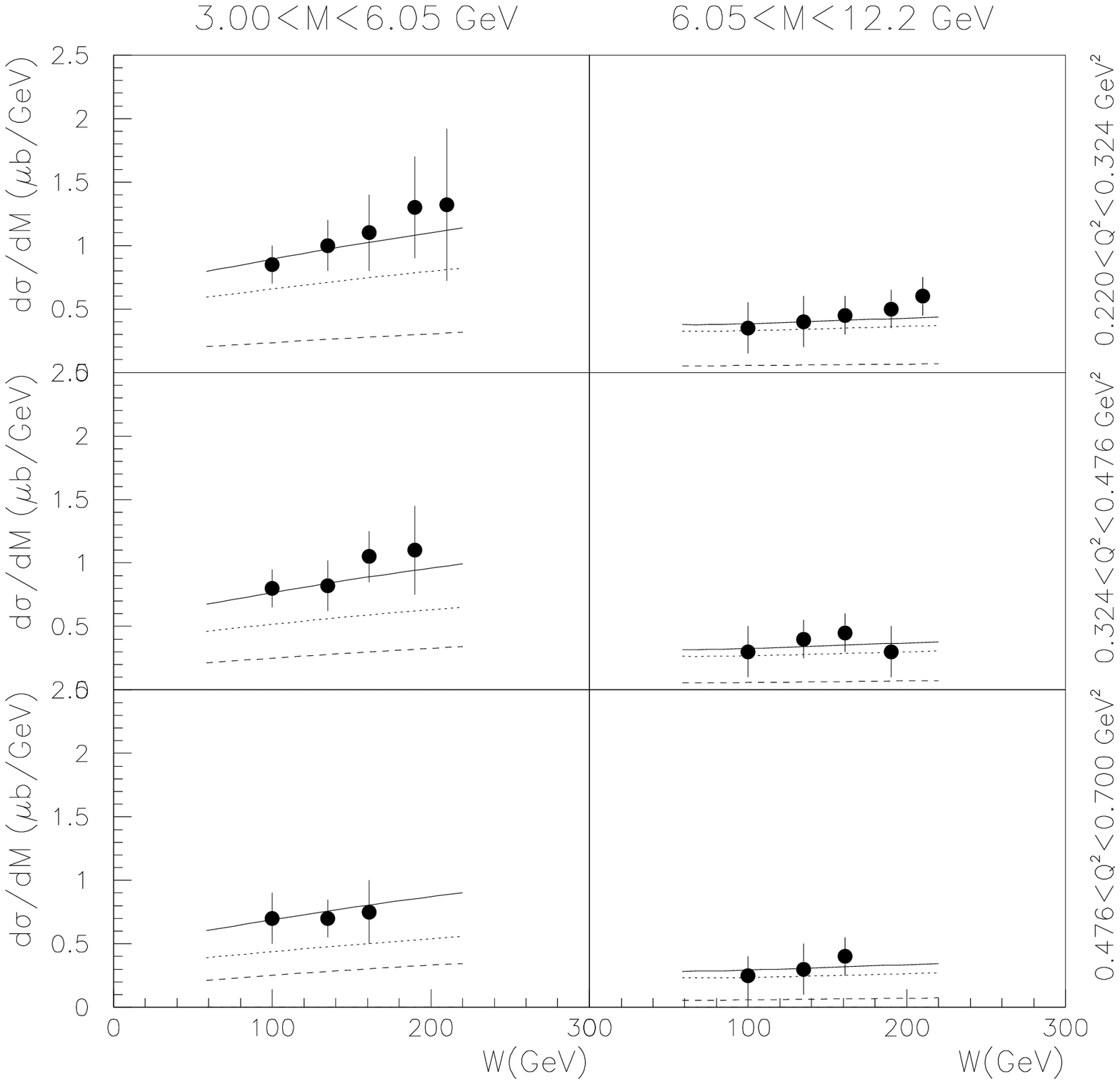,width=16.cm}
\end{center}

\newpage

\centerline{\bf Figure 6}

\vspace{1cm}

\begin{center}
\hspace{-1.2cm}\epsfig{file=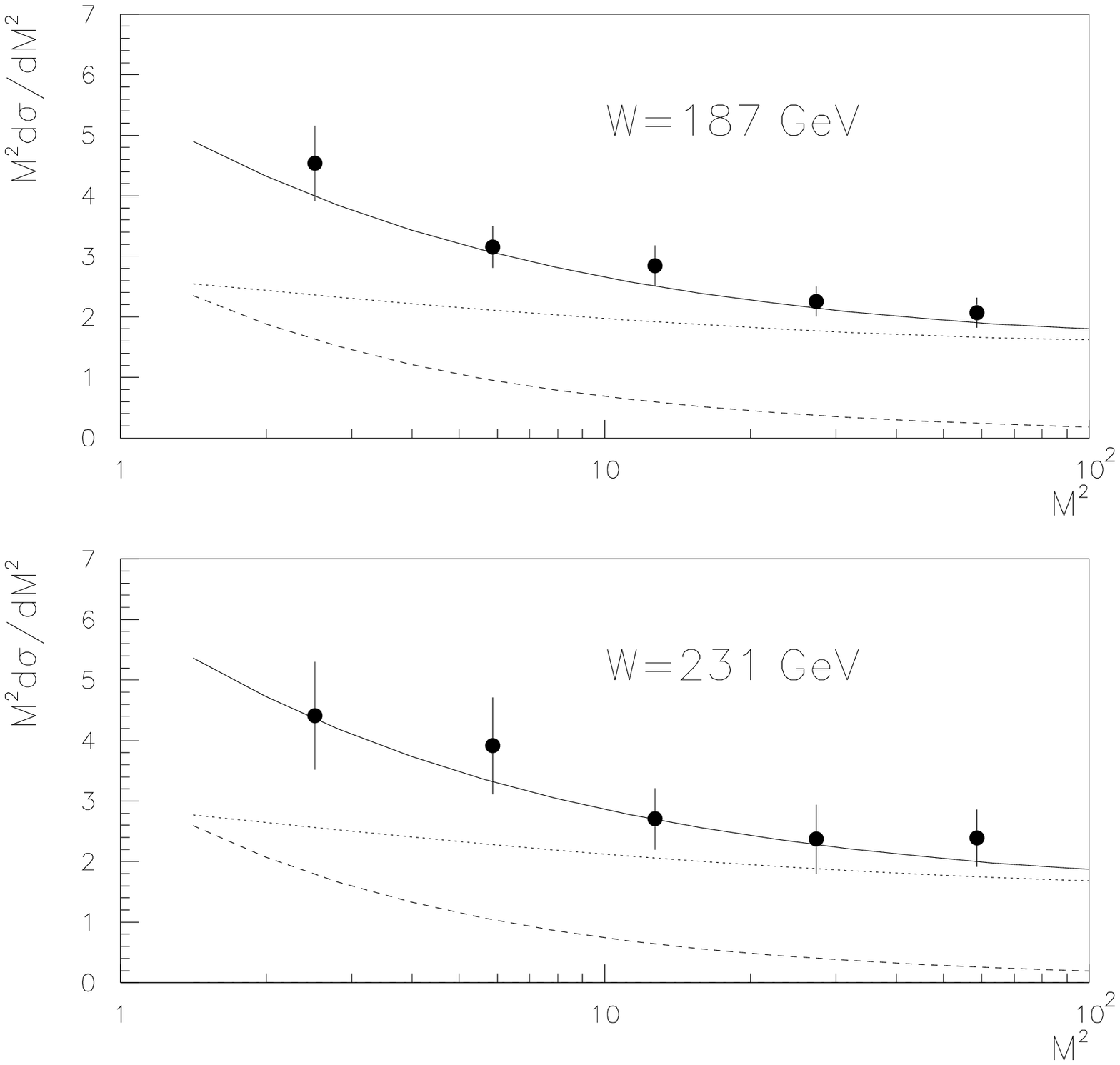,width=16.cm}
\end{center}

\end{document}